\documentclass[pra,aps,twocolumn,showpacs]{revtex4}

\usepackage{amsmath}
\usepackage{amsfonts}
\usepackage{amssymb}
\usepackage{graphicx}

\newcommand{\ME}[3]{\left\langle #1 \right| #2 \left| #3 \right\rangle}
\newcommand{\Av}[1]{\left\langle #1 \right\rangle}

\newcommand{\ket}[1]{\left| #1 \right\rangle}

\newcommand{\scalar}[2]{\left\langle #1 | #2 \right\rangle}

\newcommand{\eqr}[1]{Eq.~(\ref{#1})}
\newcommand{\fir}[1]{Fig.~\ref{#1}}

\def\be{\begin{equation}}
\def\ee{\end{equation}}
\def\bea{\begin{eqnarray}}
\def\eea{\end{eqnarray}}
\def\bes{\begin{equation*}}
\def\ees{\end{equation*}}
\def\beas{\begin{eqnarray*}}
\def\eeas{\end{eqnarray*}}

\begin{document}

\title{Dynamics of the superfluid to Mott insulator transition in one dimension}

\author{S.~R.~Clark and D.~Jaksch}
\affiliation{Clarendon Laboratory, University of Oxford, Parks
Road, Oxford OX1 3PU, United Kingdom}

\date{\today}

\begin{abstract}
We numerically study the superfluid to Mott insulator transition
for bosonic atoms in a one dimensional lattice by exploiting a
recently developed simulation method for strongly correlated
systems. We demonstrate this methods accuracy and applicability to
Bose-Hubbard model calculations by comparison with exact results
for small systems. By utilizing the efficient scaling of this
algorithm we then concentrate on systems of comparable size to
those studied in experiments and in the presence of a magnetic
trap. We investigate spatial correlations and fluctuations of the
ground state as well as the nature and speed at which the
superfluid component is built up when dynamically melting a Mott
insulating state by ramping down the lattice potential. This is
performed for slow ramping, where we find that the superfluid
builds up on a time scale consistent with single-atom hopping and
for rapid ramping where the buildup is much faster than can be
explained by this simple mechanism. Our calculations are in
remarkable agreement with the experimental results obtained by
Greiner {\em et al.} [Nature (London) {\bf 415}, 39 (2002)].
\end{abstract}


\maketitle

\section{Introduction}

Recent experiments on loading Bose-Einstein condensates into an
optical lattice have allowed for the creation and study of
strongly correlated systems of
atoms~\cite{Greiner,Esslinger,Peil,Williams}. In particular the
superfluid (SF) to Mott insulating (MI) transition first observed
in a seminal experiment by Greiner {\em et al.}~\cite{Greiner} has
received a lot of attention since it impressively demonstrated a
clean realization of the Bose-Hubbard model (BHM)~\cite{Fisher}
which has long been considered a toy model in condensed matter
physics. Furthermore, in the ideal MI state each atom is localized
to a lattice site corresponding to a commensurate filling of the
optical lattice with zero-particle-number fluctuations. These
properties make MI states attractive candidates for several
applications, most notably quantum memory, quantum
computing~\cite{Jaksch99,Bloch,Dorner,Knight,Brennen,Jaksch00,Briegel},
and quantum simulations of many-body quantum
systems~\cite{Jane,Molmer}.

The BHM Hamiltonian describes atoms loaded into a sufficiently
deep optical lattice~\cite{Jaksch98,Zwerger}. It contains a
kinetic energy term, with matrix element $J$, describing the
hopping of particles from one site to the next and an interaction
term, with matrix element $U$, which accounts for the repulsion of
two atoms occupying the same site. The ratio $U/J$ increases with
the depth of the optical lattice and can be varied over several
orders of magnitude by tuning the optical lattice
parameters~\cite{Jaksch98}. In particular by changing the
intensity of the laser beams creating the optical lattice it is
possible to vary $J$ and $U$ on time scales much smaller than the
decoherence time of the system. This opens up the possibility of
directly studying the dynamics of the BHM during the quantum phase
transition at temperature $T=0$~\cite{Greiner,Jaksch02}. According
to mean-field (MF) theory this phase transition occurs at $u_c =
U/zJ \approx 5.8$, where $z$ is the number of nearest-neighbor
sites in the lattice~\cite{Fisher,Sheshadri,Monien} and is easily
accessible in an optical lattice.

In~\cite{Greiner} the dynamics of atoms in a three dimensional
(3D) optical lattice ($z=6$) was studied while more recently
optical lattice setups where the motion of the atoms was
restricted to 1D ($z=2$)~\cite{Esslinger} were investigated. These
experiments revealed some striking properties of the quantum phase
transition. In particular a feature which is yet to be fully
understood is the time scale over which coherence is built up
throughout the atomic system when going from the MI to the SF
limit~\cite{Greiner}. Indeed it cannot be easily explained using
MF theory and numerical studies of this dynamical effect were,
until now, limited to small systems of approximately ten atoms.
Recently, however, it has been shown that quantum computations on
1D systems of qubits which do not give rise to strong entanglement
can be efficiently simulated on a classical computer via the so
called time-evolving block decimation (TEBD)
algorithm~\cite{Vidal1}. An immediate application of this
discovery is to the simulation of the time evolution of many-body
1D quantum systems which are governed by a nearest-neighbor
Hamiltonian~\cite{Vidal2}. The BHM is one of many important model
Hamiltonians which fall into this class~\cite{Daley}. The
simulation method is efficient for all such 1D model Hamiltonians
due to a universal property of 1D systems that their ground state
and lowest-lying excitations tend to contain only a small amount
of entanglement~\cite{Vidal2}.

In this paper we restrict our attention to the 1D BHM with our
physical motivation being to study the the nature and speed at
which the superfluid component is built up as the system is
dynamically driven through the SF-MI transition. By exploiting the
efficient scaling of the TEBD algorithm with the size of the
system we are able to investigate this phenomenon for setups which
are of comparable size to those studied in
experiments~\cite{Esslinger}. First, in Sec.~\ref{model}, we
introduce the 1D BHM for describing atoms in optical lattices and
briefly introduce the TEBD algorithm as used in this paper. In
Sec.~\ref{groundstates} we then demonstrate the applicability of
the TEBD to the BHM by comparison with exact numerical
calculations for small systems. This is then followed by an
investigation of SF and MI ground states of larger lattice setups
concentrating on their spatial correlations and occupation number
fluctuations together with a comparison to MF results. We then
study the dynamics of the MI to SF transition in
Sec.~\ref{dynamics} when changing the lattice depth on two
different time scales. Most notably for rapid MI to SF ramping we
find that the width of the central interference peak, as observed
after releasing the atoms from the lattice, shrinks with an
increasing total ramping time with the same functional dependence
found in~\cite{Greiner}. This result is discussed in
Sec.~\ref{dynamics}~B. Finally, we summarize our results in
Sec.~\ref{concl}.

\section{Model and Numerical Method}
\label{model}

In this section we introduce the BHM describing bosonic atoms in
an optical lattice where the motion is restricted to 1D and give a
short overview of the numerical method used in our simulations.

\subsection{Model}

By confining an ultracold bosonic gas in a 3D optical lattice with
a large depth in the two orthogonal directions $y$ and $z$ it is
possible to create an array of effective 1D systems in the $x$
direction~\cite{Esslinger,Dorner,Paredes}. The dynamics of these
systems is governed by the external trapping and the optical
lattice potential along the $x$ axis. The optical lattice then has
a depth $V_0$ proportional to the laser intensity and a lattice
period $a=\lambda /2$, where $\lambda $ is the wavelength of the
laser light. The Hamiltonian describing each 1D system reduces to
the 1D BHM (for details see Appendix \ref{1dol})~\cite{Jaksch98}
(taking $\hbar=1$ throughout):
\begin{equation}
H=\sum_{m}-J(b_m^{\dagger }b_{m+1}+{\rm H.c.})-\mu_m b_m^\dagger
b_m + \frac{U}{2} b_m^\dagger b_m^\dagger b_m b_m, \label{BHM}
\end{equation}
where the operators $b_m$ ($b_m^\dagger$) are bosonic destruction
(creation) operators for a bosonic particle in site $m$, centered
at $x_m=ma$, obeying the standard canonical commutation relations.
The grand canonical Hamiltonian then has $\mu_m = \mu - V_T(x_m)$
as the local chemical potential for site $m$, where $V_T$ is the
external trapping potential. The parameters $U$ and $J$ can be
determined in terms of the Wannier functions $w(x)$ as shown in
Appendix \ref{1dol} and under the assumptions outlined are
independent of the lattice site $m$~\cite{Jaksch98}. Their ratio
can be varied over a wide range by dynamically changing the depth
$V_0$ of the optical lattice. For all the systems considered here
we take the wavelength of the light used to form the optical
lattice as $\lambda=826$~nm and the atomic species trapped as
$^{87}$Rb, where $a_s=5.1$~nm.

\subsection{Numerical method}

In this paper we exploit the recently devised TEBD simulation
algorithm~\cite{Vidal1,Vidal2} which allows the dynamics of 1D
systems with nearest-neighbor interactions, such as the BHM, to be
computed accurately and efficiently. The TEBD algorithm has been
shown to be closely related to the density matrix renormalization
group (DMRG)~\cite{White92,White93}. Over the past decade the DMRG
has provided enormous insight into the static and dynamic
equilibrium properties of 1D systems. Although originally devised
as a ground-state method, it has been extended to yield accurate
low-energy spectra~\cite{Jeckelmann}, and also to calculations of
the real time evolution of 1D systems~\cite{Cazalilla} which is of
particular importance here. The approach of~\cite{Cazalilla} is to
take the DMRG ground state $\ket{\psi(0)}$ obtained for the
initial Hamiltonian and use it to define a decimation of the
Hilbert space in which the Schr{\"o}dinger equation is numerically
integrated. The key assumption, and most severe approximation,
within this scheme is that this static subspace defined by
$\ket{\psi(0)}$ is adequate to approximate $\ket{\psi(t)}$ with
reasonable accuracy for all times. In general this will only be
true for short periods of time. Novel methods have been
devised~\cite{Wang} which can maintain the accuracy over longer
periods by `targeting' other states in addition to the ground
state, but in doing so the efficiency of the computation is
significantly reduced~\cite{White04}. In contrast the TEBD
algorithm can maintain typical DMRG accuracies while remaining
efficient. Despite their differing origins it has recently been
shown that TEBD and DMRG algorithms share some crucial conceptual
and formal similarities~\cite{Daley,White04}. Indeed both methods
search for an approximation to the true wave function within a
restricted class of wave functions which are described by matrix
product states~\cite{Fannes,Rommer}, and do so with identical
decomposition and truncation procedures. The essential difference,
which we shall emphasize shortly, is that the TEBD algorithm
updates the matrix product decomposition directly and in such a
way that the resulting decimated subspace in which the time
evolution is computed is optimally adapted at each
step~\cite{Daley}.

Here we briefly outline the essential features of the TEBD
algorithm, with specific attention to its application to the BHM.
Let us consider a 1D BHM composed of $M$ sites. An arbitrary state
of this system can be expanded in the Fock basis
\begin{equation}
\ket{\psi}=\sum_{n_1=0}^{\infty}\cdots\sum_{n_M=0}^{\infty}c_{n_1
\cdots n_M}\ket{n_1,\dots,n_M}, \label{expand}
\end{equation}
where $\ket{n_m}$ denotes the Fock state of $n_m$ particles in
site $m$. For the purpose of simulating this system the number of
Fock basis states per lattice site must be cut off to some upper
limit $n_{\textrm{max}}$. In all the numerical calculations we
performed $n_{\textrm{max}}=5$. This is sufficient to avoid any
cut off effects in the bosonic occupation, as long as only small
filling factors of the lattice are used and the on-site
interaction energy $U$ is sufficiently large compared to the
hopping energy $J$.

Now suppose we split the system into two contiguous parts $A_m$
composed of the first $m$ sites and $B_m$ composed of the last
$M-m$ sites. We can think of this partitioning as cutting the $m$
th bond situated between sites $m$ and $m+1$. For any state
$\ket{\psi}$ a Schmidt decomposition (SD) can be performed which
renders the state in the form
\begin{equation}
\ket{\psi}=\sum_{\alpha=1}^{\chi_m}\lambda_{\alpha}^{[m]}\ket{\phi_{\alpha}^{A_m}}\ket{\phi_{\alpha}^{B_m}},
\end{equation}
where $\chi_m$ is the Schmidt rank of the SD,
$\lambda_{\alpha}^{[m]}$ are the Schmidt coefficients, and
$\ket{\phi_{\alpha}^{c}}$, with $c \in \{ A_m, B_m \}$, are the
corresponding Schmidt states of the respective subsystems. The
Schmidt rank $\chi_m$ is a useful measure of the entanglement
between the two subsystems $A_m$ and $B_m$~\cite{Vidal1}. Given
any state $\ket{\psi}$ a set of $(M-1)$ SD can be performed
according to a sequence of such partitions of the system with $m
\in \{1\cdots M-1\}$, as depicted in \fir{chidiagrams}(a).

\begin{figure}[Htp]
\begin{center}
\includegraphics[width=8.5cm]{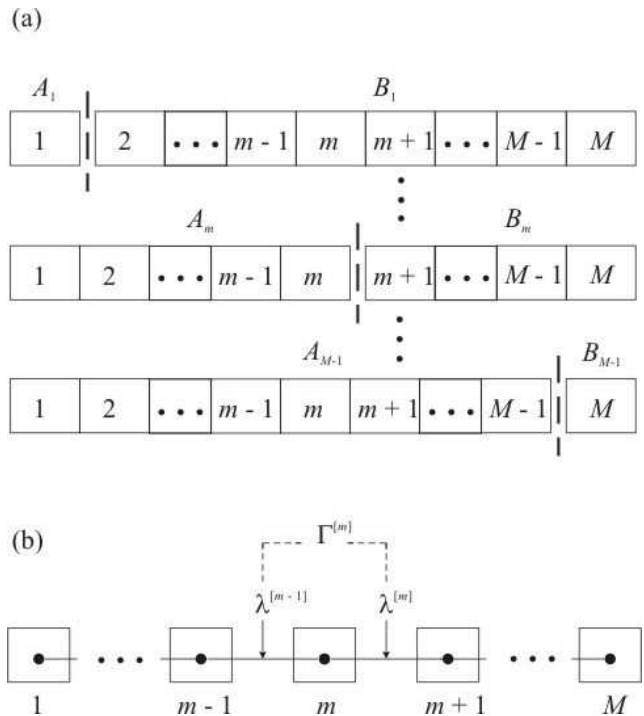}
\caption{(a) The sequence of contiguous partitions of the system
in which the SD are computed. The coefficients and states from
these SD are then used to form the $\Gamma$ and $\lambda$ tensors.
(b) A depiction of the $\Gamma$ tensors associated to lattice
sites and $\lambda$ tensors associated to bonds between those
sites.} \label{chidiagrams}
\end{center}
\end{figure}

Using the $\lambda_{\alpha}^{[m]}$ and states
$\ket{\phi_{\alpha}^{s}}$ for each subsystem obtained from these
SD it is possible~\cite{Vidal1} to construct a set of $\Gamma$ and
$\lambda$ tensors which are equivalent to a matrix product
decomposition of the expansion coefficients $c_{n_1 \cdots n_M}$
of $\ket{\psi}$ in the fixed Fock basis~\cite{Daley}. Specifically
one finds
\begin{equation}
c_{n_1 \cdots n_M}=\sum_{\alpha_1, \dots ,\alpha_{M-1}}
\Gamma_{\alpha_1}^{[1]n_1}\lambda_{\alpha_1}^{[1]}\Gamma_{\alpha_1
\alpha_2}^{[2]n_2}\lambda_{\alpha_2}^{[2]} \cdots
\lambda_{\alpha_{M-1}}^{[M-1]} \Gamma_{\alpha_M}^{[1]n_M},
\label{chiexp}
\end{equation}
where $n_m$ is the occupation number of site $m$, and $\alpha_m$
are the Schmidt indices of the $m$ th partition, each of which
sums from 1 to its respective Schmidt rank $\chi_m$. With
reference to \fir{chidiagrams}(b) we note that each
$\lambda_{\alpha_{m}}^{[m]}$ is labeled by the bond between sites
$m$ and $m+1$, along with the corresponding Schmidt index
$\alpha_m$, whereas each $\Gamma_{\alpha_{m-1}
\alpha_{m}}^{[m]n_m}$ is labeled by a site $m$ which resides
between the two bonds $m-1$ and $m$, and so also possess the
Schmidt indices $\alpha_{m-1}$ and $\alpha_m$ of these bonds.

Under the circumstances described the expansion, \eqr{chiexp}, is
exact and as such the number of parameters stored could grow
exponentially with the size of the system. However, it is a
general feature of 1D systems with nearest-neighbor interactions
that the entanglement within their ground state and low-lying
excitations depends weakly on the size of the
system~\cite{Vidal2}. Indeed it can be shown that the entanglement
of a block of size $\ell$ with the rest of the system remains
finite as $\ell \rightarrow \infty$ in 1D systems or at worst
grows logarithmically with $\ell$ at criticality~\cite{Daley}.
Consequently the entanglement between the blocks of any of the
$(M-1)$ SD illustrated in \fir{chidiagrams}(a) can be saturated by
some fixed Schmidt rank, which for the systems we consider is
typically small. It is this fact that accounts for the success of
DMRG in 1D systems. Similarly within TEBD it allows the maximum
possible Schmidt rank used in the matrix product decomposition,
\eqr{chiexp}, to be fixed to some value $\chi$, thereby truncating
it to the most significant contributions. For an appropriate
choice of $\chi$ this approximation will be accurate with the
error proportional to the sum of the discarded eigenvalues in the
SD~\cite{Vidal2}. This clear interpretation of the central
numerical parameter $\chi$ within TEBD is very useful. Once a
value of $\chi$ is found to saturate the entanglement of the
ground state and low-lying excitations of a system then this a
direct measure of the role of entanglement in the dynamics of the
system. In total the scaling in the number of parameters within
the expansion, \eqr{chiexp}, is quadratic in $\chi$ and linear in
the size of the system $M$ and in $n_{\textrm{max}}$. So upon
fixing $\chi$ and thus preventing its possible exponential
dependence on $M$, the description becomes efficient. As with DMRG
this decomposition of a state generates, for all practical
purposes, an optimal $\chi \times \chi$ matrix product
state~\cite{Daley}. A noteworthy limit of this is the
approximation is where $\chi=1$, which forces the description of
the system to be of product form with respects to all sites. Using
the TEBD algorithm under this severe restriction is in fact
equivalent to MF theory and the Gutzwiller ansatz
\cite{Fisher,Sheshadri,Krauth,Stoof,Jaksch02}.

Another crucial advantage of the TEBD algorithm is that once a
state is expressed in the matrix product form, \eqr{chiexp}, one-
and two-site unitary transformations can be applied directly and
exactly to the system such that the resulting state can be
efficiently returned to a matrix product form~\cite{Daley}.
Indeed, given a partitioning of the system into a
two-block-two-site configuration $[1 \cdots m-1][m \quad m+1][m+2
\cdots M]$, the application of a two-site unitary to sites $m$ and
$m+1$ only requires updates to be performed on the tensors {\it
local} to those sites--namely, $\Gamma_{\alpha_{m-1}
\alpha_m}^{[m]n_m}\lambda_{\alpha_m}^{[m]}\Gamma_{\alpha_m
\alpha_{m+1}}^{[m+1]n_{m+1}}$. The major computational effort of
this update is limited to the rediagonalization of the reduced
density matrix of one of the adjacent site and block subsystems,
such as sites $[m+1][m+2 \cdots M]$, which is of dimension $(\chi
n_{\textrm{max}}) \times (\chi n_{\textrm{max}})$ at
most~\cite{Vidal1}. The crucial feature here is that a DMRG-style
truncation to only the most relevant eigenstates of this reduced
density matrix occurs in an optimal way at each application of a
two-site unitary. This is in contrast with time-dependent DMRG
methods where the basis states which make up the matrix product
decomposition are fixed at the start~\cite{Cazililla,Daley}. The
number of basic operations required to perform this update scales
as {\em O}$(\chi^3)$~\cite{Vidal1}.

To compute the action of the time-evolution unitary $\exp{(-iH
\delta t)}$, for a time step $\delta t$, we first make the
observation that for Hamiltonians with nearest-neighbor
interactions, which are composed of two-site operators at most,
terms can be separated into a sum of those involving odd sites,
$F$, and those involving even sites, $G$:
\begin{eqnarray}
F & = & \sum_{n \textrm{ odd}}{F_{n,n+1}}, \\
G & = & \sum_{n \textrm{ even}}{G_{n,n+1}},\\
H & = & F + G.
\end{eqnarray}
Given that no terms within $F$ involve the same lattice sites they
all commute amongst themselves. Thus the action of
$\exp{(-iF\delta t)}$ can be computed exactly as
\begin{equation}
e^{-iF \delta t} = \prod_{n \textrm{ odd}}{e^{-i F_{n,n+1} \delta
t}}.
\end{equation}
Since each term in this product is a two-site unitary, they can be
applied individually to the state with the method detailed
in~\cite{Vidal1}, and the same is also true for $G$. The
complications in computing the time evolution arise from the fact
that $F$ and $G$ do not in general commute, and hence we
approximate the unitary time-evolution operator $\exp(i(F+G)\delta
t)$ using a Trotter expansion. Ignoring their noncommutativity
would constitute a first-order expansion. If we define
\begin{equation}
s_2(F,G,y) = e^{-iFy/2}e^{-iGy}e^{-iFy/2},
\end{equation}
then the second-order expansion follows when $y=\delta t$. For the
numerical simulations performed in this paper the fourth-order
expansion \cite{Suzuki} was used, which has the form
\begin{equation}
 e^{-i(F+G) \delta t} = \prod_{l=1}^{5} s_2(F,G,q_l \delta t) + \textrm{\em O}(\delta
 t^5),
\end{equation}
where the parameters $q_l$ are defined as
\begin{equation}
q_1 = q_2 = q_4 = q_5 \equiv q = \frac{1}{(4-4^{1/3})},~q_3 =
1-4q.
\end{equation}
A detailed analysis of the errors and computational cost of TEBD
is given in~\cite{Vidal2}, where it is shown that the Trotter
error propagates quadratically with the simulated time and so the
accuracy of the method can be maintained for long periods with
appropriate choices of the parameters.

The pure TEBD implementation we employ here can be improved
further by combining the advantageous features of TEBD outlined
with the well-established optimizations of DMRG such as good
quantum numbers and White's `state prediction' method. In doing so
an adaptive time-dependent DMRG algorithm is
obtained~\cite{Daley,White04} illustrating the extremely close
relationship between these two methods. Finally we note the very
recent advances in generalizing TEBD and DMRG to describe mixed
state dynamics and generic master equation evolution of 1D systems
with nearest-neighbor coupling~\cite{Vidal3,Cirac}. This opens up
the possibility of simulating finite temperature effects,
decoherence and dissipation.

\section{Ground states of the BHM}

\label{groundstates} We first investigate the ground state of the
BHM and compare the numerical results with the exact ground states
for a small homogeneous system. Then we consider a larger system
in the presence of a shallow magnetic trap $V_T$ superimposed on
the lattice and compare the results to those predicted by MF
theory. In all cases the numerical ground state was computed with
the TEBD algorithm using continuous imaginary time evolution from
a simple product state, as detailed in \cite{Vidal2}.

\subsection{Comparison of exact and simulated ground states}

To investigate the accuracy of the numerical simulation and its
applicability to the BHM we first consider a small system in which
an exact solution can be found readily. Specifically we use an
optical lattice composed of $M=7$ sites, a trapping potential of
$V_T=0$ with box boundary conditions, and a total number of
particles $N=7$. The ground state is then calculated numerically
and exactly for $U/2J=2,~6$ and $20$, corresponding to the SF,
intermediate, and MI regimes, respectively. The numerical
simulation was performed for $\chi=3,~5$ and $7$ in each case.

The one-particle density matrices $\rho_{m,n}=\langle
b_{m}^\dagger b_n \rangle$ obtained for each regime for the
numerical and exact calculations are visually indistinguishable in
all cases. In order to highlight the extent of the agreement we
present a number of other plots. Specifically in the SF regime the
comparisons of the spatial correlation of the central site
$|\rho_{4,4+d}|$ as a function of the distance $d$ are shown in
\fir{plots_2}(a) and \fir{plots_2}(b) between the exact and
numerical calculations using $\chi=3$ and $\chi=5$. Identical
comparisons of the standard deviation of the site occupation
$\sigma(\rho_{m,m}) = (\langle N_m^2 \rangle - \langle N_m
\rangle^2)^{1/2}$, where $N_m=b_m^{\dagger}b_m$, and the spectrum
of the one-particle density matrices $e_\gamma$ [normalized with
tr$(\rho)=N$] are shown in \fir{plots_2}(c), \fir{plots_2}(d) and
\fir{plots_2}(e),\fir{plots_2}(f) respectively. These results show
that although there is qualitative agreement between exact and
$\chi=3$ calculation, almost all expectation values have a maximum
deviation from the exact calculation improved by an order of
magnitude with $\chi=5$. As expected for a SF ground state the
one-particle density matrix spectrum in \fir{plots_2}(f) is
dominated by one eigenvalue of order $N$. However, given that the
lattice still has a nonzero depth this state deviates from that of
a pure SF, where $\ket{\psi_{\textrm{SF}}} \propto
(\sum_m{b_m^{\dagger }})^N \ket{\textrm{vac}}$, since the sum of
the remaining eigenvalues (in descending order) is
$\sum_{\gamma=2}^7 e_{\gamma} \approx 2.5$ and so represents a
significant quantum depletion of the SF.

For the intermediate and MI regime a similar factor of improvement
can be obtained; however in this case the $\chi=3$ calculation
already yields excellent agreement with the exact calculation.
Specifically we find the infidelity between the numerical and
exact many-body state is $1-F=1-|\scalar{\psi_0}{\psi_0'}| \approx
10^{-5}$, where $\ket{\psi_0}$ ($\ket{\psi_0'}$) is the numerical
(exact) ground state and the temperature corresponding to the
difference in their ground-state energy as $\Delta T \approx
10^{-2}$ nK. The comparisons of the spatial correlation, site
occupancy standard deviation, and one-particle density matrix
spectrum for $U/2J=6$ and $U/2J=20$, with $\chi=3$, are shown in
\fir{plots_6_20}.

These plots, along with those where $\chi=5$ for the SF regime,
illustrate the onset of increasing MI characteristics in the
ground states. In particular the rapidly decreasing spatial
correlations and fluctuations in occupancy, as well as the change
in the spectrum from being dominated by one single-particle state
to having seven almost equally occupied orbitals. These indicate
that the MI ground state obtained is a very close approximation to
that of a pure MI, where $\ket{\psi_{\textrm{MI}}} \propto
\prod_m{b_m^{\dagger }} \ket{\textrm{vac}}$, representing
commensurate filling of the lattice. However given that the
lattice is not infinitely deep deviations with this pure MI state
exist and are evident from the persistence of small off-diagonal
correlations visible at $d=1$ in \fir{plots_6_20}(b) and the
spread of the spectrum about unity in \fir{plots_6_20}(f).

As expected we find that the agreement between the numerical and
exact calculations for a given value of $\chi$ improves with
increasing $U/2J$ in line with the decrease in off-diagonal
correlations. In all cases the $\chi=7$ results gave excellent
agreement with the exact calculation. The worst case being in the
SF regime where an infidelity of $1-F \approx 10^{-4}$, and a
deviation in ground-state energy of $\Delta T \approx 10^{-2}$ nK
was obtained.

We note here that the exact calculation was performed with the
canonical Hamiltonian with $N=7$, whereas the numerical
simulations used the grand canonical Hamiltonian. With an
appropriate choice of the chemical potential $\mu$ the average
particle number can be fixed to $N=7$, enabling the comparisons
above. Indeed for the calculations performed the worst absolute
value of the projection of the simulated state outside the $N=7$
Fock subspace was $\ME{\psi_0}{\mathcal{P}_{N\neq7}}{\psi_0}
\approx 10^{-13}$. Hence our results confirm the agreement of
these approaches for small systems, and we assume the agreement
holds for the larger systems.

\begin{figure}[h!]
\begin{center}
\includegraphics[width=8.5cm]{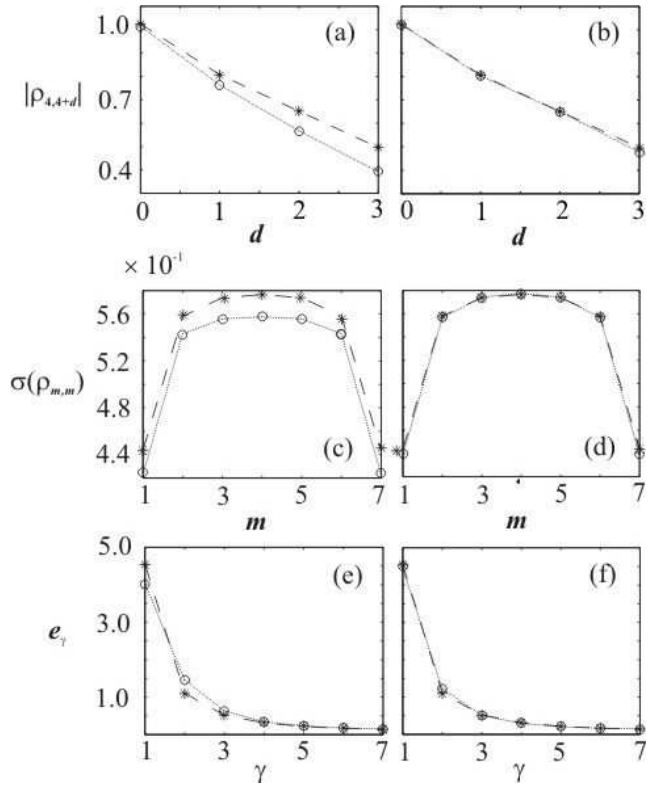}
\caption{Comparisons of the numerical (o) and exact (*)
calculations with $U/2J=2$ for spatial correlations
$|\rho_{4,4+d}|$ with the central site $m=4$ obtained for (a)
$\chi = 3$ and (b) $\chi = 5$, the standard deviation of the site
occupation $\sigma(\rho_{m,m})$ obtained for (c) $\chi = 3$ and
(d) $\chi = 5$, and the spectrum $e_{\gamma}$ of the one-particle
density matrix obtained for (e) $\chi = 3$ and (f) $\chi = 5$. The
dashed and dotted curves shown are to guide the eye.}
\label{plots_2}
\end{center}
\end{figure}

\begin{figure}[h!]
\begin{center}
\includegraphics[width=8cm]{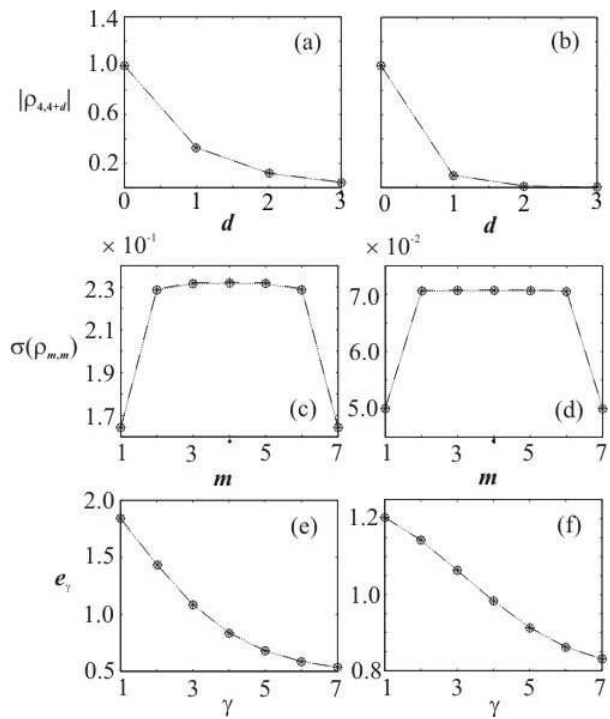}
\caption{Comparisons of the numerical (o) and exact (*)
calculations, where the numerics were all performed with $\chi =
3$, for spatial correlations $|\rho_{4,4+d}|$ with the central
site $m=4$ obtained with (a) $U/2J=6$ and (b) $U/2J=20$, the
standard deviation of the site occupation $\sigma(\rho_{m,m})$
obtained with (c) $U/2J=6$ and (d) $U/2J=20$, and the spectrum
$e_{\gamma}$ of the one-particle density matrix obtained with (e)
$U/2J=6$ and (f) $U/2J=20$. Note the differing scales and that the
dashed and dotted curves are shown only to guide the eye.}
\label{plots_6_20}
\end{center}
\end{figure}

\subsection{MI and SF states with a superimposed magnetic trap}

To consider systems closer to those studied in experiments
\cite{Esslinger} we use a lattice with $M=49$ sites and made the
system inhomogeneous by superimposing a harmonic trap potential
$V_T(x_m)=m_A \omega^2 x_m^2 /2$, where $\omega$ is the trapping
frequency and $m_A$ is the mass of an atom. As with the smaller
system the ground states for the SF, intermediate and MI regimes
were calculated. The lattice was loaded with a total number of
particles $N=40$ by choosing an appropriate chemical potential
$\mu$ in each regime. For all cases a trapping frequency of
$\omega=2\pi\times 70$ Hz was used and found to be sufficient in
eliminating any occupation at the boundaries of the system.

The inhomogeneity caused by a spatially varying confining
potential can result in the coexistence of spatially separated SF
and MI regions. Such properties have been confirmed
experimentally~\cite{Greiner,Esslinger} and have received intense
theoretical study with numerical calculations. In particular
through Gutzwiller ansatz and MF theory~\cite{Jaksch98,Bruder} and
Quantum Monte Carlo (QMC)~\cite{Batrouni,Bergkvist} and
DMRG~\cite{Kollath} simulations in 1D, as well as calculations
using the QMC~\cite{Wessel,Kashurnikov} and numerical
renormalization group~\cite{Pollet} methods for 2D and 3D systems.
Here we explore the SF-MI coexistence features of BHM ground
states in order to confirm the physical picture arising from our
numerical calculation for the large inhomogeneous system.
Specifically we make comparisons of the mean site occupancy and
its standard deviation between the numerical results and those
obtained by MF theory for each regime (see Appendix \ref{mfapp}
for details of the MF calculation)~\cite{Bruder,Stoof}.

\begin{figure}[h!]
\begin{center}
\includegraphics[width=9cm]{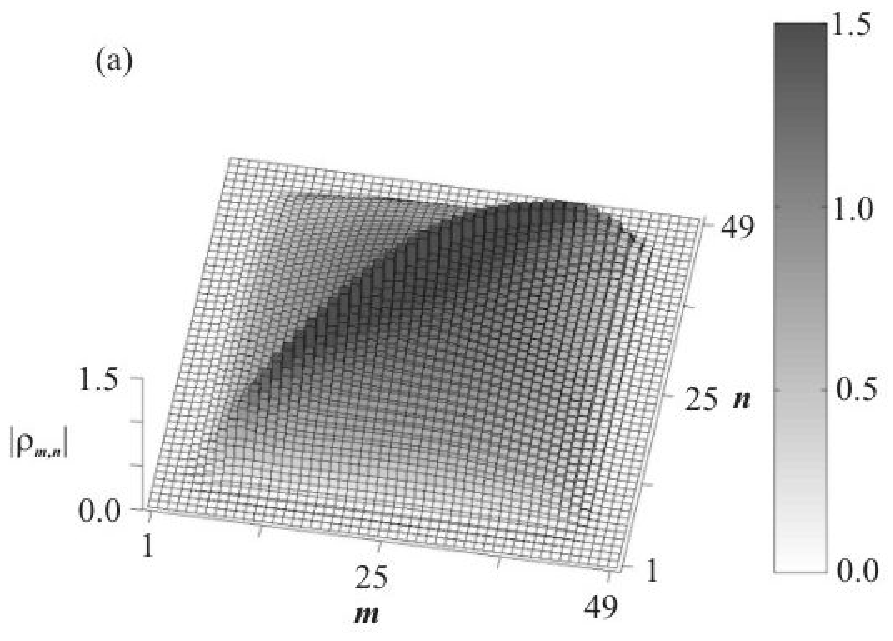}
\includegraphics[width=9cm]{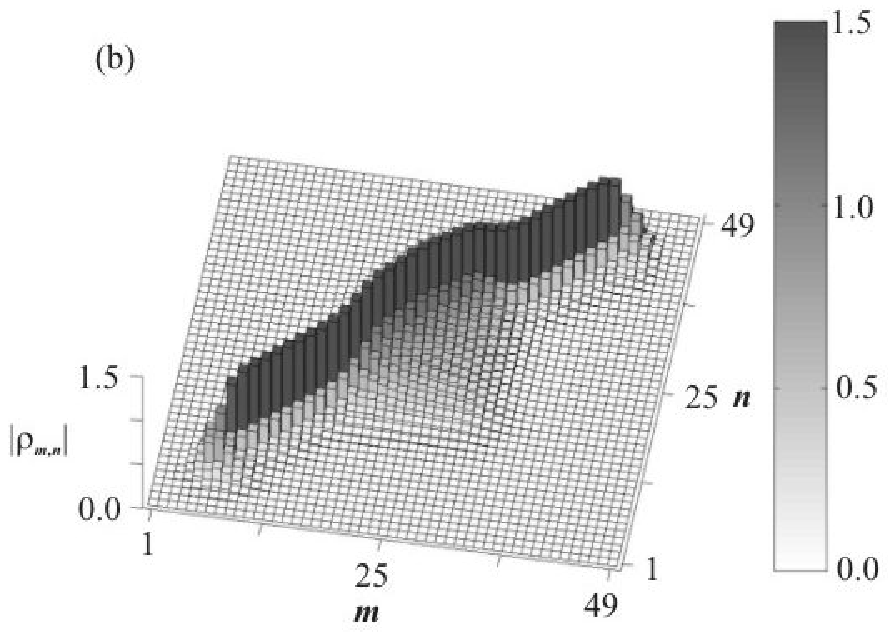}
\includegraphics[width=9cm]{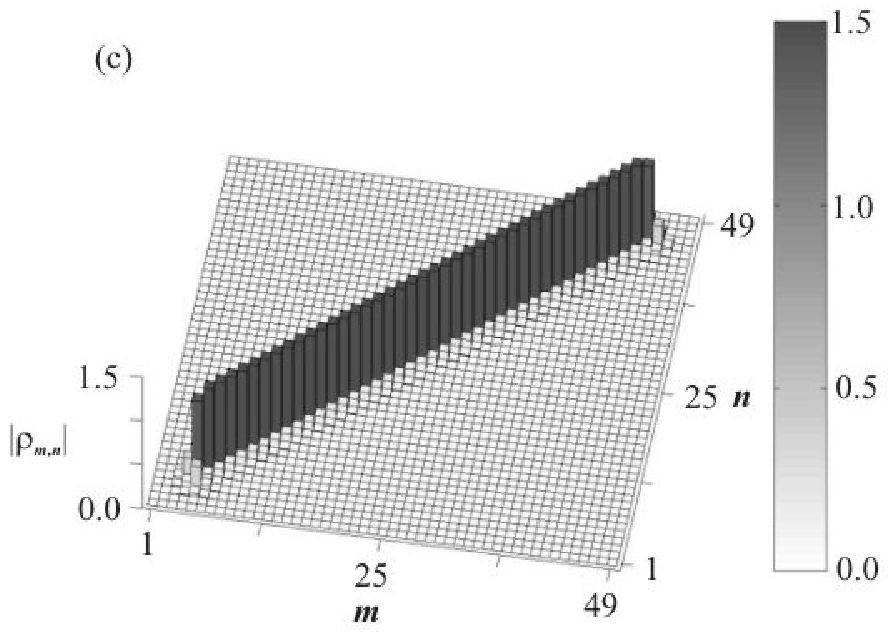}
\caption{The absolute value $|\rho_{m,n}|$ of the one-particle
density matrix as a function of site indices $m,n$ for (a)
$U/2J=2$, (b) $U/2J=6$, and (c) $U/2J=20$.} \label{sprdm_large}
\end{center}
\end{figure}

\begin{figure}[h!]
\begin{center}
\includegraphics[width=8.5cm]{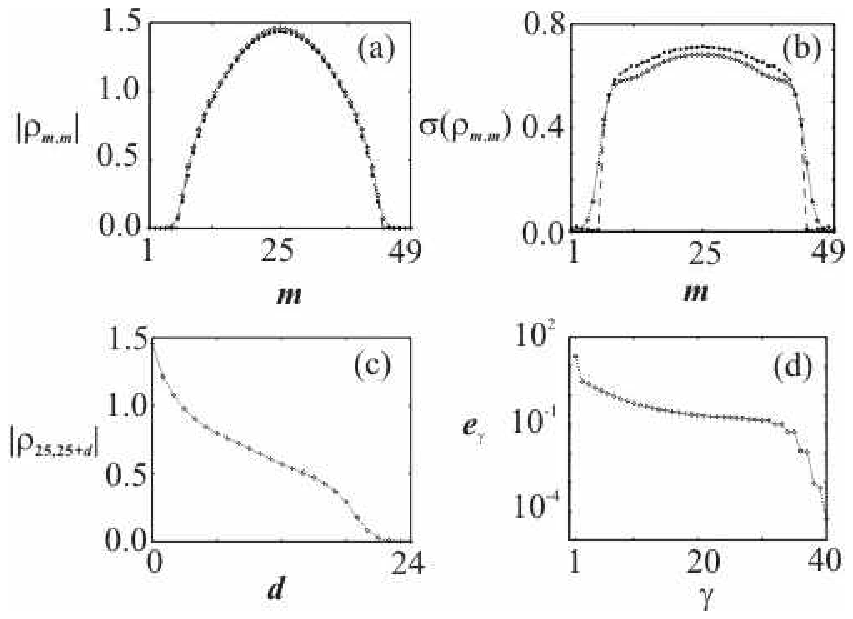}
\includegraphics[width=8.5cm]{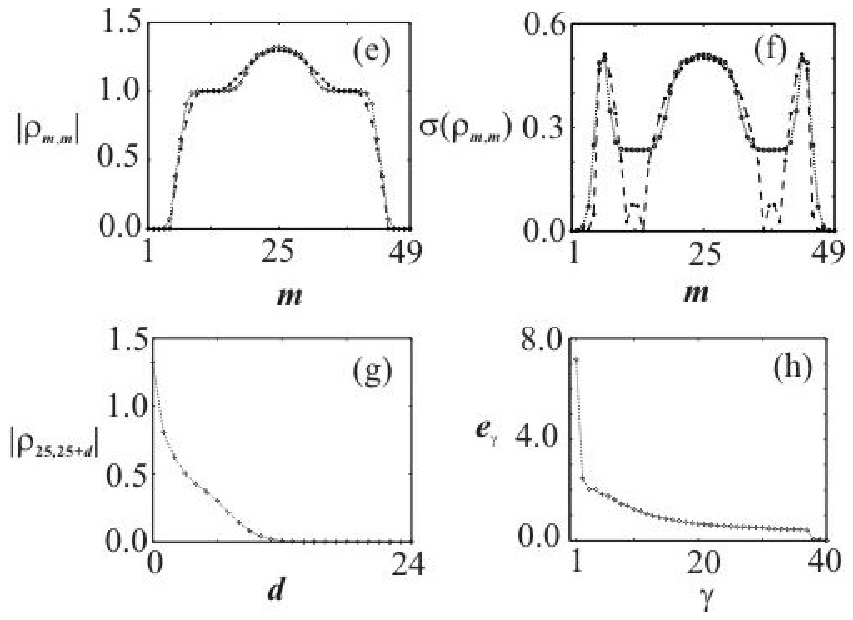}
\includegraphics[width=8.5cm]{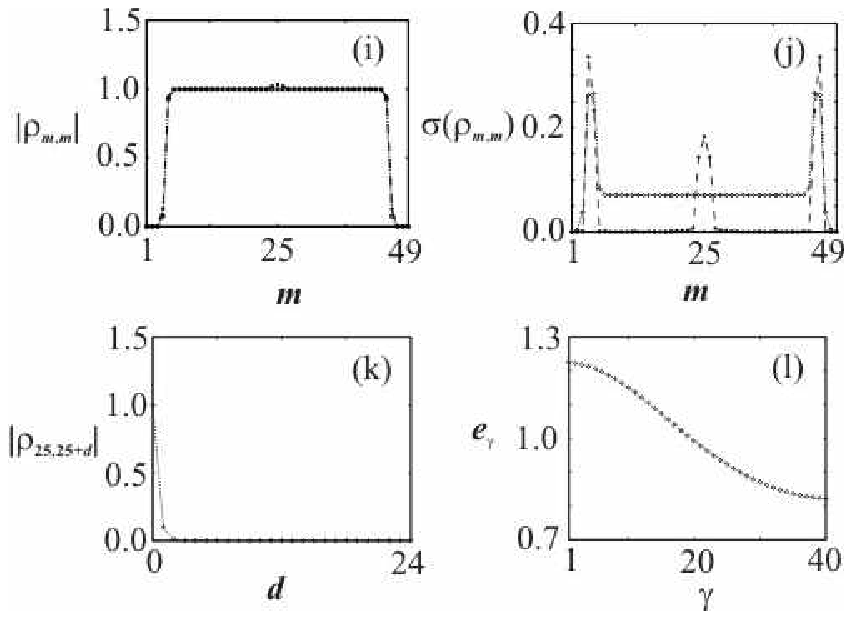}
\caption{Specific plots for the three regimes. For $U/2J=2$ there
is (a) the site occupancy $|\rho_{m,m}|$, (b) the standard
deviation of the site occupancy $\sigma(\rho_{m,m})$, both with
the MF calculation shown as the dashed curve, (c) the spatial
correlations from the central site $m=25$, and (d) the spectrum
$e_{\gamma}$ of the one-particle density matrix, showing only the
40 nonzero eigenvalues. The same plots are presented for $U/2J=6$
in (e)-(h) and for $U/2J=20$ in (i)-(l).} \label{plots_large}
\end{center}
\end{figure}

The one-particle density matrix of the resulting SF ground state
is shown in \fir{sprdm_large}(a). Important features of this state
are outlined in \fir{plots_large}(a)-\fir{plots_large}(d), where
it can be seen that the system is entirely SF. The MF results for
the site occupancy and its standard deviation, are also shown in
\fir{plots_large}(a) and \fir{plots_large}(b), and as expected
there is good agreement between the curves in both cases. For the
intermediate regime, whose one-particle density matrix is shown in
\fir{sprdm_large}(b), we see a system possessing alternating
regions of coherent SF and incoherent MI phases
\cite{Kollath,Batrouni}. The pattern of these regions starts with
a central SF region with a mean occupancy exceeding unity, which
then becomes a singly occupied MI, a SF region with mean occupancy
less than unity, and finally the vacuum MI. In
\fir{plots_large}(e) and \fir{plots_large}(f) we see that the MI
region in \fir{plots_large}(e) coincides with suppressed
fluctuations in occupancy shown in \fir{plots_large}(f). The MF
curves also plotted show general agreement with these phase
identifications; however the MF curve in \fir{plots_large}(f)
predicts a significantly greater suppression of the particle
number fluctuations for the MI regions than the numerical results.
Such deviations are consistent with the fact that MF theory
predicts a sharp and well-pronounced SF-MI phase transition
\cite{Bruder,Stoof}. While these prediction are known to be
accurate for infinite homogeneous systems, for small inhomogeneous
systems we see that the role of correlations is important and that
the transition between the SF and MI regions is not established
with such definiteness.

Given that the ground state for the intermediate regime exhibits
significantly sized SF regions which are separated by a MI, we
investigated whether any correlations were present between these
regions. Evidence of such correlations would be the presence of
elongated peaks in the one-particle density matrix at the
off-diagonal locations corresponding to the intersection of the
rows $m$ and columns $n$ of the SF regions. To within the accuracy
obtained with $\chi=5$ no such correlation peaks were found in the
one-particle density matrix of the ground state. This was
confirmed with continuous imaginary time evolution not only for a
product initial state used conventionally, but also with an
initial state which already contained significant correlations.
Our results suggest that if such SF correlations do exist within
the ground state, then they are extremely fragile. Despite this we
shall see shortly that such correlations do occur readily in
dynamical situations which cross the MI-SF transition, where the
system does not necessarily remain in the ground state.

Lastly, the one-particle density matrix for the MI regime is shown
in \fir{sprdm_large}(c). It is clear from this and the
corresponding plots in \fir{plots_large}(i) and
\fir{plots_large}(j) that the system is almost completely in the
singly occupied MI phase, aside from the small SF regions at the
far extremes before the vacuum. Their presence is typified by the
two peaks in the occupancy standard deviation shown in
\fir{plots_large}(j). The MF calculation in this case gives the
same identification of the regions, except for the very center of
the trap, where a small SF region is predicted to exist, as
evidenced by the central peak in the MF occupancy standard
deviation curve of \fir{plots_large}(j). Again MF theory predicts
a much greater suppression of occupancy fluctuations in the MI
regions of the system than the numerical calculation. As with the
smaller system the plots of \fir{plots_large} for the three ground
states illustrate the transition from predominately SF to MI
characteristics.

In all cases the ground-state calculations were performed with
$\chi=5$. To ensure convergence the calculations were repeated for
$\chi=7$. The largest deviation between these two calculations was
found in the SF regime where the estimated ground-state energy
differed by a temperature $\Delta T_{\chi=5\rightarrow7} \approx
0.2$~nK. We made a similar comparison between the $\chi=5$ results
and those of MF theory, which are equivalent to $\chi=1$, where
the largest deviation was found to be $\Delta
T_{\chi=1\rightarrow5} \approx 3$~nK. Given the larger occupancy
of the system the calculations were also repeated with larger
values of $n_{\textrm{max}}$, confirming that no cut off effects
were encountered.

\section{Dynamics of the BHM}
\label{dynamics}

The most novel feature of the TEBD algorithm is its capacity to
efficiently simulate the dynamics of 1D systems which are
inaccessible to exact calculation. Here we consider dynamics which
are generated by varying the optical lattice depth $V_0(t)$
according to some ramping profile in time. The results of this is
that the parameters $J(t)$ and $U(t)$ in the BHM Hamiltonian of
\eqr{BHM} becoming time dependent. By appropriately choosing the
range of values covered by the optical depth $V_0(t)$ it is
possible to dynamically drive the system through the SF-MI
transition. We shall consider such dynamics occurring on two
different time scales: first via a slow and smooth profile, and
second as fast linear ramping. Our objective being in both cases
to observe the nature and speed in which coherence is
reestablished within the system.

\subsection{Slow dynamics}

\subsubsection{Profile of the slow dynamics}

For the slow dynamical profile a smoothed `box' function was used
for the depth $V_0(t)$. Such a profile for the dynamics has been
considered before for small systems \cite{Jaksch02}. It has the
form
\begin{equation}
V_0(t) = V_{\textrm{SF}} + \mathcal{N}\frac{V_{\textrm{MI}} -
V_{\textrm{SF}}}{1+e^{[(t-t_c)^2-t_w^2]/t_s^2}}, \label{box}
\end{equation}
where $t_c$, $t_w$, and $t_s$ are time parameters specifying the
center, width, and step size of the profile respectively, while
$\mathcal{N}=1+e^{-t_w^2/t_s^2}$ is the scaling factor required to
ensure that the depth varies from $V_{\textrm{SF}}$ to
$V_{\textrm{MI}}$. The lattice depths $V_{\textrm{SF}}$ and
$V_{\textrm{MI}}$ were chosen to be the depths equivalent to
$U/2J=2$ and $U/2J=20$, respectively, in correspondence with the
parameters used in the previous section for the SF and MI regime.
The exact shape of the profile for $U/2J$ resulting from the
ramping of $V_0(t)$ chosen is shown in \fir{small_dyn}(a). Large
time parameters have been used in order to keep the time evolution
of the system sufficiently adiabatic and to prevent excessive
excitations.

\subsubsection{Slow dynamics of the small system: $M=7$}

First we consider the slow dynamics applied to the small system.
This provides the opportunity to solve the BHM dynamics both
numerically and exactly, allowing a direct comparison of the
accuracy and a demonstration of the applicability of the algorithm
to the dynamics of the BHM. The system was initially prepared in
the SF ground state computed earlier in Sec.~\ref{groundstates}~A,
and using $\chi=5$ for the numerical calculations. The time
evolution was then performed for a total time
$t_{\textrm{tot}}=2t_c$, with time $t$ running over the interval
$[0,t_{\textrm{tot}}]$. The spectrum $e_{\gamma}$ of the
one-particle density matrix $\rho_{m,n}(t)$ is plotted as a
function of time in \fir{small_dyn}(b). For times $t$ where
$U/2J<u_c$ the spectrum is, as expected, dominated by one large
eigenvalue whose value is of order of the number of atoms. As
$U/2J$ crosses the MF critical value $u_c$ the eigenvalues are
found to converge around the region of 1. Indeed the state of the
system given by the numerical calculation at the time $t=t_c$ in
the dynamics, which corresponds to $U/2J=20$, is found to have an
infidelity with the numerical MI ground state computed earlier in
Sec.~\ref{groundstates}~A as $1-F<10^{-4}$. This confirms that for
a small and homogeneous system the ramping is sufficiently
adiabatic to ensure that the system has entered the MI regime as
the ground state and the one-particle density matrix is diagonal.

With decreasing optical depth and, in turn, decreasing $U/2J$, the
SF ground state is restored when $U/2J=2$ is reached again at $t=2
t_c$. The infidelity between the initial numerical SF ground state
and the final numerical SF state was found to be $1-F < 10^{-3}$.
In \fir{small_dyn}(c) the behavior of the fluctuations in the site
occupancy is as expected; namely, the standard deviation in site
occupancy is suppressed with increasing lattice depth, and
restored with its subsequent decrease.

To test the accuracy of the TEBD algorithm a number of comparisons
to the exact calculation were made. The simplest of these was the
maximum relative deviation between the exact and numerical results
for the one-particle density matrix spectrum $\Delta e =
\textrm{max}_{\gamma}(|1-e_{\gamma}/e_{\gamma}'|)$, where
$e_{\gamma}$ and $e_{\gamma}'$ are the numerical and exact
results, respectively. The time profile $\Delta e$ is plotted in
\fir{small_dyn}(d). It is found that over the whole time evolution
the relative deviation is at most $\Delta e \approx 10^{-1}$. A
similar relative deviation can be defined for the standard
deviation of the occupancy as $\Delta \sigma =
\textrm{max}_{m}(|1-\sigma_{m}/ \sigma_{m}'|)$, where it is found
that $\Delta \sigma \approx 10^{-2}$ at most during the time
evolution.

The most conclusive comparison, however, is the infidelity
$1-F(t)$ of the exact and numerical many-body states over the time
evolution, shown in \fir{small_dyn}(e). It is clear from this that
the infidelity is bounded as $1-F(t)< 4 \times 10^{-3}$ over the
whole evolution. The shape of the infidelity profile also gives
important information about the TEBD method. Namely, it fits the
general observation made in Sec.~\ref{groundstates}~A for the
ground-state calculations that for fixed numerical parameters
$\chi$ and $n_{\textrm{max}}$ the simulation is more accurate in
the intermediate and MI regimes than the SF regime. This behavior
is precisely exhibited in the time dependence of the above
comparisons where significant reductions in the deviations are
seen when the system enters the MI regime.

\begin{figure}[h!]
\begin{center}
\includegraphics[width=9cm]{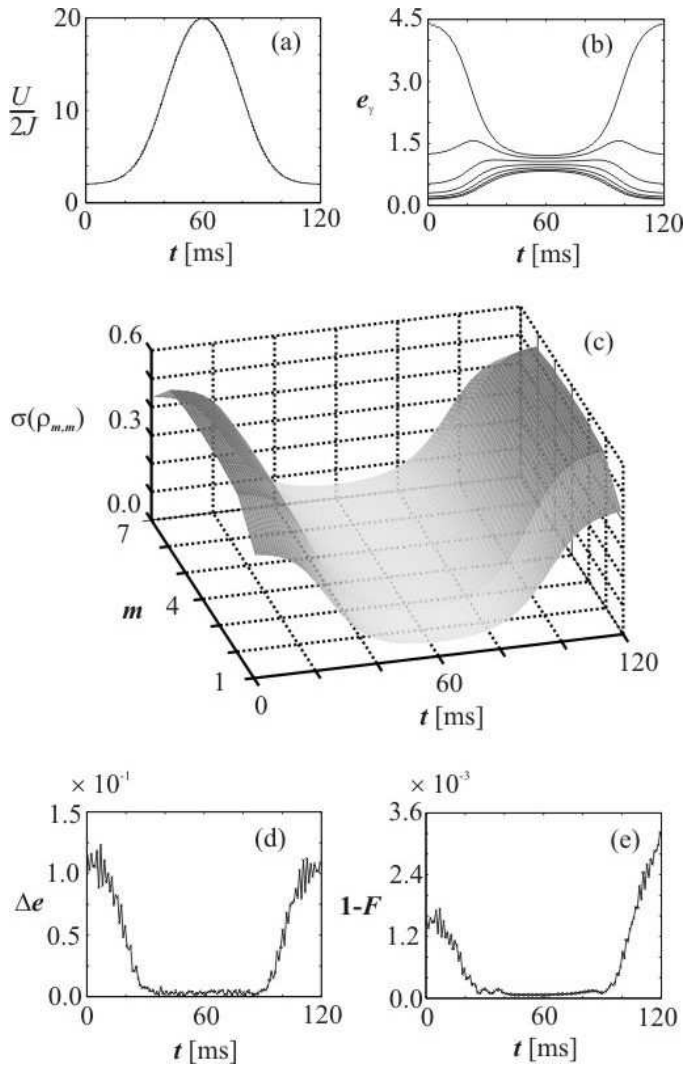}
\caption{Slow dynamics of the small system: (a) the resulting
ramping profile of the parameter $U/2J$ with time, where the time
parameters for the $V_0(t)$ profile are $t_c=60$~ms, $t_w=24$~ms,
and $t_s=18$~ms, (b) the spectrum of the one-particle density
matrix $e_{\gamma}$ with time obtained from the numerical
calculation, (c) the standard deviation of the site occupancy
$\sigma(\rho_{m,m})$ with time obtained from the numerical
calculation, (d) the maximum deviation between the numerical and
exact spectrum $\Delta e$ with time, and (e) the infidelity $1-F$
of the numerical many-body state compared to the exact state with
time.} \label{small_dyn}
\end{center}
\end{figure}

\subsubsection{Slow dynamics of the larger system: $M=49$}

The larger system possesses many of the essential characteristics
of current experimental implementations of the SF-MI
transition~\cite{Greiner,Esslinger}. In particular these include
the inhomogeneous nature of the system, caused by a trapping
potential, and the larger number of both lattice sites, and atoms,
as compared with the smaller system. With a linear size of $M=49$
sites the system considered is on the same scale as experiments
already performed~\cite{Greiner,Esslinger}. The major difference
is that the mean occupancy at the center of our system is
$\Av{n_c} \approx 1.5$, roughly half that of most experiments,
where it is usual to have $\Av{n_c} \approx 2.5$. While the mean
occupancy undoubtedly has an important influence in the dynamics,
the system simulated here is sufficiently close that it can
demonstrate much of the important physics.

The slow ramping profile, \fir{small_dyn}(a), was performed
identically on the larger inhomogeneous system using the ground
state \fir{sprdm_large}(a) computed earlier as the initial state
and for a total time $t_{\textrm{tot}}=3t_c$. The resulting
spectrum $e_{\gamma}$ of the one-particle density matrix is
plotted as a function of time in \fir{slow_dyn}. General features
of this spectrum follow from the smaller homogeneous system;
namely, the trend for the eigenvalues to decrease as the bottom of
the ramping is reached. While most eigenvalues converge to unity,
as with the smaller system, one in particular can be seen to
remain much larger than this during the entire dynamics. This is a
clear indicator of a significant SF region within the state as it
is dynamically driven into the MI regime. For the larger
inhomogeneous system we see that the slow ramping profile is not
adiabatic enough to bring the system into the MI ground state
shown in \fir{sprdm_large}(c).

Indeed to examine the nature of the state generated by the
dynamics additional plots of the one-particle density matrix are
also shown in \fir{slow_dyn} for the two times indicated during
the dynamics. These show that the state of the system remains
close in form to that of the ground state for the intermediate
regime shown in \fir{sprdm_large}(b), where a large SF region
exists at the center of the trap. However, unlike that ground
state we see that sizable correlations between the separated SF
regions, which were alluded to earlier in
Sec.~\ref{groundstates}~B, do exist and remain present even at the
bottom of the profile for \fir{slow_dyn}(ii) where $U/2J=20$.

Another important difference between the spectra of the small and
larger system is the more prominent excitations which have been
induced during the transition. These are visible as the
oscillatory behavior of the eigenvalues seen in the latter section
of the profile in \fir{slow_dyn}. Their presence is consistent
with the fact that larger systems have more numerous and closely
spaced low-lying excitations, however, despite this the
oscillations have only a small amplitude and so do not destroy the
SF obtained at the end of the transition.

\begin{figure}[h!]
\begin{center}
\includegraphics[width=9cm]{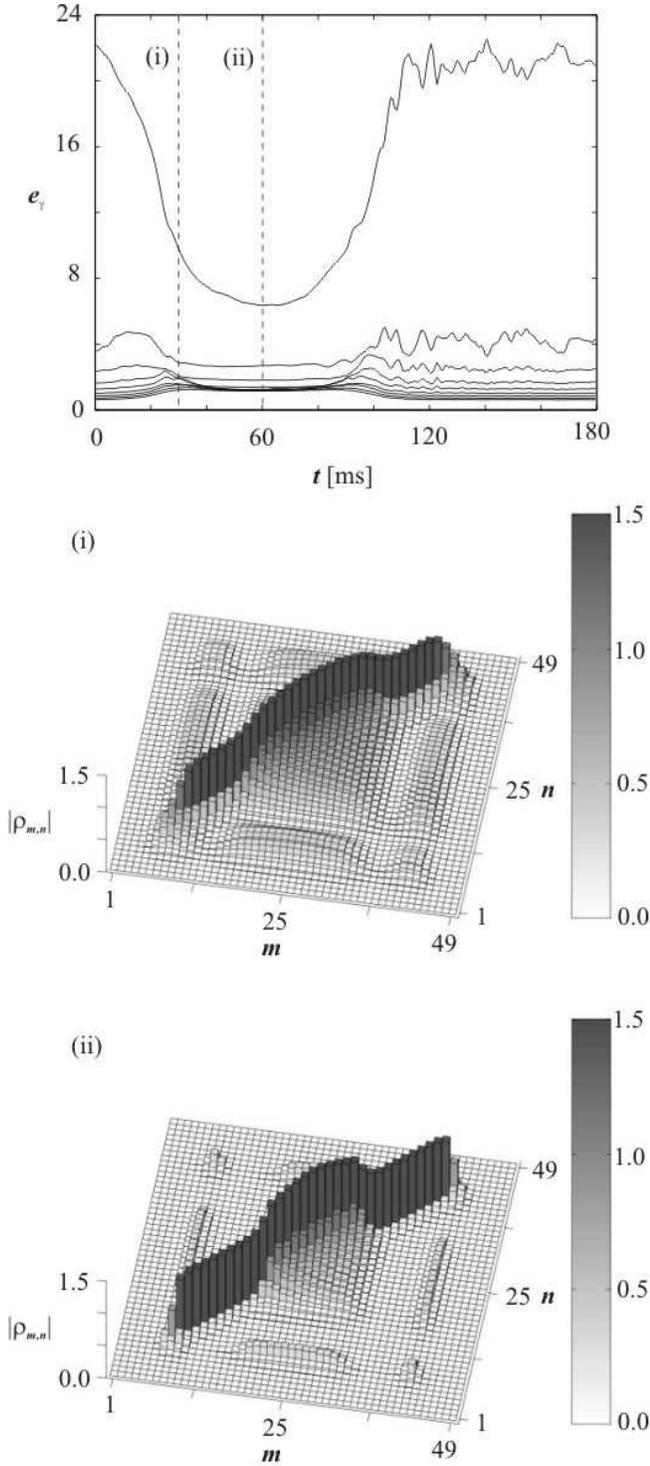}
\caption{The spectrum of the one-particle density matrix
$e_{\gamma}$ for the slow dynamics of the larger inhomogeneous
system, showing only the largest ten eigenvalues. The dashed lines
denote the two times (i) $t=t_c/2$ and (ii) $t=t_c$ for which the
absolute value $|\rho_{m,n}|$ of the one-particle density matrix
is plotted.} \label{slow_dyn}
\end{center}
\end{figure}

In order to examine the speed at which coherence is reestablished
in the system during the latter half $(t>t_c)$ of the ramping
profile the correlation length of the system must be computed over
time. The correlation length is typically defined as the distance
at which the off-diagonal elements of the one-particle density
matrix become negligible \cite{Stringari}. For symmetrical
systems, like those considered here, it is natural to measure this
from the central site $m=25$. However, the inhomogeneity of the
system, which results in the kind correlations between spatially
separated SF regions just discussed for \fir{slow_dyn}(i), makes
the determination of the correlation length ambiguous. Instead we
choose to examine a cutoff length $\xi_c$ where the spatial
correlations with the central site have a specific value
$|\rho_{25,25+\xi_c}|=1/e\approx 0.37$. This value is large enough
that it corresponds to tracking a point on the central SF region
and so can provide a relative measure of its size. The change in
$\xi_c$ over time is plotted in \fir{slow_corr_fft}(a). The same
plot also shows the fitted curve whose function is that of a
smooth `well', which is the reflection of the smooth `box' used
earlier in \eqr{box} about the $t$ axis. The time parameters of
this fit are very close to those of the resulting `well' for
$J(t)$ generated from the $V_0(t)$ profile. The variation in
$\xi_c$ over time demonstrates that it is capturing the essential
changes in the central SF region, including the oscillations in
its size at later times caused by excitations.

\begin{figure}[h!]
\begin{center}
\includegraphics[width=8.8cm]{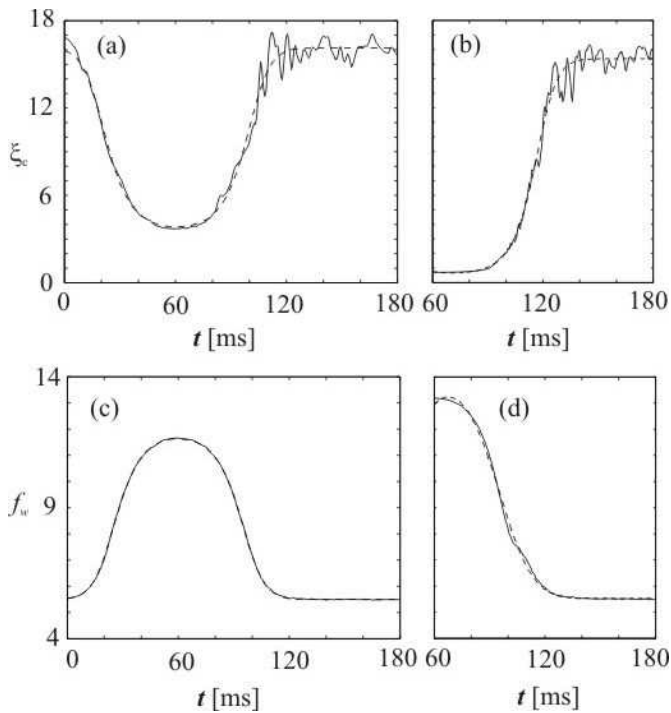}
\caption{Slow dynamics of the larger inhomogeneous system with the
variation in the correlation cutoff $\xi_c$, measured in lattice
sites, for (a) the complete slow ramping profile starting from the
SF ground state, and for (b) the MI ground state beginning at the
bottom of the ramping. The momentum distribution width $f_w$, also
measured in lattice sites, for the same situations is shown in (c)
and (d) respectively. The dashed curves present in all plots are
the fitted smooth `box' or `well' functions to the data.}
\label{slow_corr_fft}
\end{center}
\end{figure}

To investigate how much the presence of residual SF correlations
effects the time dependence of $\xi_c$ the latter half ($t>t_c$)
of the same slow ramping profile was applied to the MI ground
state of \fir{sprdm_large}(c) found earlier for $U/2J=20$. In
contrast to the state which is dynamically driven to the MI regime
this ground state has virtually no off-diagonal correlations for
any site. The change in $\xi_c$ of time for this case is shown in
\fir{slow_corr_fft}(b), and as before the fitted smooth `well'
function is also plotted. These show that despite $\xi_c$ starting
from a much smaller value for the MI ground state it still
acquires the mean value of the final SF state on the same time
scale as that of the dynamically driven state.

An important and experimentally motivated measure of the coherence
of a state can be obtained from its momentum distribution function
$p_k$. In experiments the interference pattern resulting from the
state of the system is examined by allowing all the atoms within
the lattice to expand freely for a short period of time and then
measure the absorption $I(x)$ at points $x$ on a distant
observation line. In the simplest model of this process one can
neglect both the interactions between atoms during the expansion
and the spatial dependence of the interference caused by the
freely evolved Wannier function envelopes
$w(x-x_m)$~\cite{Burnett}. This gives the generic features of the
interference pattern at an observation point $x$ in terms of the
path phases acquired by each site in a 1D lattice of phase
coherent matter wave sources. In the far-field approximation the
intensity $I(x)$ along the observation line is proportional to the
momentum distribution $p_k \propto \sum_{m,n} \exp[ik(m-n)]
\rho_{m,n}$~\cite{Burnett,Kashurnikov}.

The form of the momentum distribution is sufficiently well behaved
that its width $f_w$ can be determined most easily by taking its
standard deviation. The variation of $f_w$ in time for the full
dynamics is shown in \fir{slow_corr_fft}(c), along with the fitted
smooth `box' function. The time parameters of this fit are again
very similar to those of $J(t)$. As expected it is seen that $f_w$
increases in line with the decrease in off-diagonal correlations.
The time profile of $f_w$ for the half-ramping of the MI ground
state is shown in \fir{slow_corr_fft}(d) and again confirms that
the momentum distribution width of the SF is reestablished on
approximately the same time scale as that of the dynamically
driven state.

Finally we examine the speed at which the correlation cutoff
length $\xi_c$ increases with time over the latter half of the
slow ramping. At any given time $t$ the characteristic time scale
at which single-atom hopping occurs is given by
$\tau_{\textrm{tunnel}}(t)=\pi/2J(t)$~\cite{timescale}. The
simplest description of the growth of the central SF region is
based on atoms at the edge of the system hopping towards the
center. In this way correlations can be established over the whole
lattice of $M$ sites~\cite{Greiner}. An estimate for the overall
time scale for this mechanism to occur is given by
$t_{\textrm{restore}}=M \tau_{\textrm{tunnel}}/2$~\cite{jchoice},
which for the system and depths used here has a value
$t_{\textrm{restore}} \approx 23$~ms. In \fir{corr_buildup} the
speed $\partial \xi_c /\partial t$ obtained from the function
fitting is plotted for (i) the full dynamics of the ramping and
(ii) the half-ramping from the MI ground state. In line with the
plots of $\xi_c$ in \fir{slow_corr_fft}(a) and
\fir{slow_corr_fft}(b) we see that there is a time delay before
there is a significant rate of change in $\xi_c$ for the MI ground
state ramping. Since the restoration of correlations occurs over
the same total time scale in both cases the peak in the
correlation speed is higher for the MI ground state. In addition
to these curves the characteristic tunnelling speed
$v_{\textrm{tunnel}}(t)=1/\tau_{\textrm{tunnel}}(t)$ over the
ramping is also shown, and most importantly we note that neither
of the two correlation speeds (i) nor (ii) exceeds this curve.
This confirms that the ramping applied is sufficiently slow that
the propagation of the SF is dominated by single atom hopping.

\begin{figure}[h!]
\begin{center}
\includegraphics[width=8.5cm]{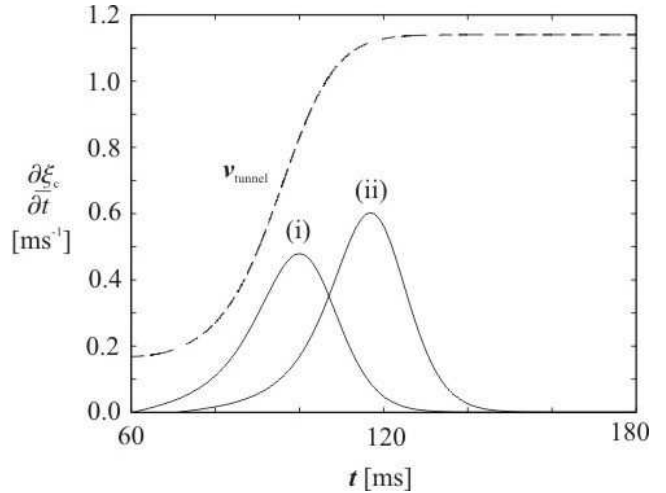}
\caption{A comparison between the speeds at which the correlation
cutoff length $\xi_c$ changes in time for (i) the full slow
ramping profile starting from the SF ground state and (ii) the
last half of the ramping profile starting from the MI ground state
at the bottom of the ramping. The time profile for the single-atom
hopping speed $v_{\textrm{tunnel}}(t)$ is also shown. All speeds
are expressed in lattice sites per ms.} \label{corr_buildup}
\end{center}
\end{figure}

\subsection{Fast dynamics}
\label{fastdyn}

\subsubsection{Profile for the fast dynamics}

The time scale over which the slow ramping occurs is of the order
of 60~ms and so greater than $t_{\textrm{restore}}$. Here we
consider ramping occurring much more rapidly. Specifically we
replace the latter part ($t>t_c$) of the slow ramping profile with
a linear ramping of the optical depth $V_0(t)$ from
$V_{\textrm{MI}}$ to $V_{\textrm{SF}}$ as
\begin{equation}
V_0(t) = V_{\textrm{MI}} - \frac{(V_{\textrm{MI}}-
V_{\textrm{SF}})}{t_{\textrm{ramp}}}(t-t_c), \label{linear}
\end{equation}
where $t$ runs from $t_c$ to $t_c + t_{\textrm{ramp}}$, and the
total time of the ramping is $t_{\textrm{ramp}}$. This gives a
total ramping profile similar to that studied experimentally by
Greiner {\em et al}~\cite{Greiner}.

\subsubsection{Fast dynamics of the larger system: $M=49$}

For the fast dynamics we restrict our attention to the state that
is dynamically driven to the MI regime by the slow ramping profile
at $t=t_c$. A number of simulation runs were performed for total
ramping times $t_{\textrm{ramp}}$ between 0.1~ms and 10~ms, along
with $t_{\textrm{ramp}}=0$~ms which is equivalent to the initial
state. The value of $\xi_c$ obtained at the end of each of the
ramping times is plotted in \fir{rapid_dyn}(a). We see that there
is a steady monotonic increase in the $\xi_c$ for ramping times
$t_{\textrm{ramp}}$, except where it is broken by peaks and
troughs which are the expected manifestations of the trapping
used. In particular the trough centered around $t_{\textrm{ramp}}
\approx 7$~ms corresponds to the period of an oscillation with
frequency $2\omega$, where $\omega$ is the trapping frequency
introduced in Sec.~\ref{groundstates}~B. On a similar basis the
spikes which appear around $t_{\textrm{ramp}} \approx 1$~ms can be
seen to be a result of the excitation spectrum.

\begin{figure}[h!]
\begin{center}
\includegraphics[width=8.5cm]{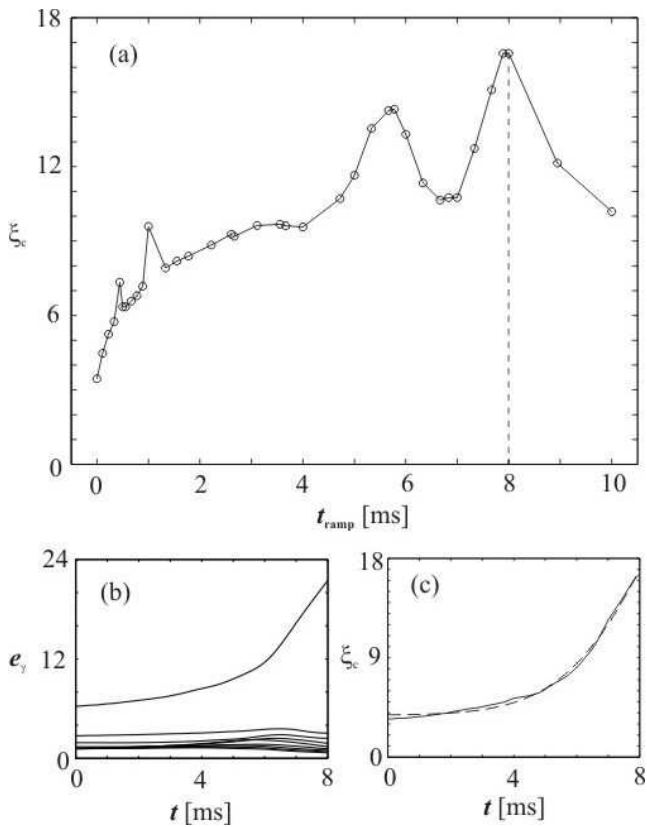}
\caption{Results for rapid dynamics, (a) the final correlation
cutoff length $\xi_c$ obtained for different linear ramping times
$t_{\textrm{ramp}}$, focusing on $t_{\textrm{ramp}}=8$ ms we have
(b) the spectrum of the one-particle density matrix $e_{\gamma}$
showing only the largest ten eigenvalues, and (c) the variation in
$\xi_c$ over the simulation run, along with a fitted smooth `box'
function shown as the dashed curve.} \label{rapid_dyn}
\end{center}
\end{figure}

We take a special interest in the ramping time
$t_{\textrm{ramp}}=8$~ms where $\xi_c$ obtains its maximum value
approximately equal to that of the SF ground state. The variation
in time of the spectrum $e_{\gamma}$ of the one-particle density
matrix and $\xi_c$ during this particular ramping simulation is
given in \fir{rapid_dyn}(b) and \fir{rapid_dyn}(c). A well-behaved
monotonic increase in $\xi_c$ is observed which can be accurately
fitted over the interval [0,8]~ms by a smooth `box' function, as
used earlier. This again provides the basis for computing the
speed $\partial \xi_c / \partial t$ at which the correlation
cutoff length $\xi_c$ is increasing over the ramping and is shown
in \fir{main_result}(b) along with that of the characteristic
tunnelling speed $v_{\textrm{tunnel}}(t)$ for the rapid ramping
profile. Unlike the similar comparison for the slow dynamics we
see here that after approximately $3$~ms $\xi_c$ is increasing in
time much more rapidly than the single-site tunneling speed
$v_{\textrm{tunnel}}$ alone can account for. Indeed by the end of
the ramping $\partial \xi_c / \partial t$ is almost three times
that of the maximum tunneling speed. This is a clear indication
that single atom hopping is not adequate to describe the growth of
the central SF region for such rapid dynamics. Instead the
specific form and contributions of higher order correlation
functions must play a crucial role.

\begin{figure}[ht]
\begin{center}
\includegraphics[width=8.5cm]{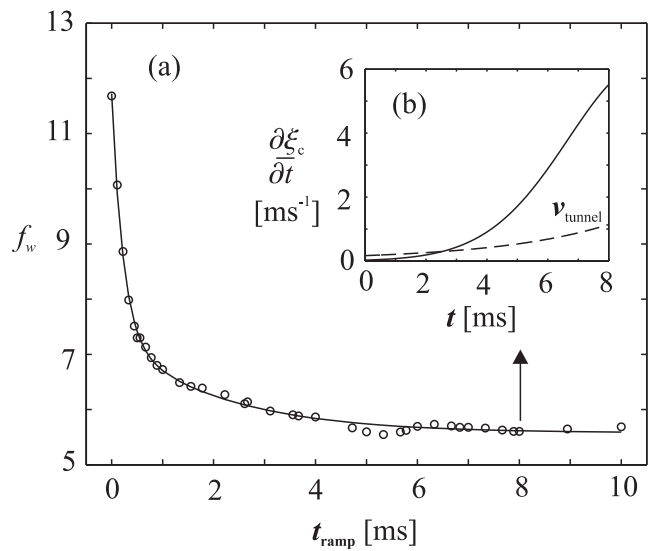}
\caption{Rapid ramping of an optical lattice from the MI
($U/2J=20$) to the SF ($U/2J=2$) regime for $N=40$ atoms in $M=49$
lattice sites superimposed by a magnetic trapping potential. The
width of the central interference fringe $f_w$ as a function of
the ramping time $t_{\textrm{ramp}}$ is shown in (a). The solid
curve is a fit using a double-exponential decay ($\tau_1 = 0.22$
ms, $\tau_2 = 2.14$ ms) (cf.~\cite{Greiner}). In (b) the rate of
change of the correlation cutoff length $\xi_c$ is shown for the
ramping performed with $t_{\textrm{ramp}}=8$ ms, along with the
profile for the characteristic tunneling speed
$v_{\textrm{tunnel}}(t)$ for the ramping. Both are plotted in
units of lattice sites per ms.} \label{main_result}
\end{center}
\end{figure}

To draw direct comparisons with the results of Greiner {\em et
al.}~\cite{Greiner} on the restoration of coherence we also plot
the momentum distribution width $f_w$ obtained from the
one-particle density matrix at the end of each ramping in
\fir{main_result}(a). The data points for this quantity show a
pronounced trend, without any of the large-scale variations seen
in $\xi_c$ caused by the trapping~\cite{Kashurnikov}. The decrease
of $f_w$ with increasing $t_{\textrm{ramp}}$ fits well to a
double-exponential decay curve of the form
\begin{equation}
f_w(t_{\textrm{ramp}}) = A_1~e^{-t_{\textrm{ramp}}/\tau_1} +
A_2~e^{-t_{\textrm{ramp}}/\tau_2} + C, \label{dblexp}
\end{equation}
where $\tau_1,\tau_2$ are the characteristic decay times,
$A_1,A_2$ are coefficients, and $C$ is a constant. Most notably
the exact same functional form was found to fit the experimental
data in~\cite{Greiner}. Since their experiment was conducted for a
3D lattice, along with a larger mean occupancy and a deeper
ramping profile, exact agreement for the time parameters of this
fit is not expected. However, we do note that the ratio of the
decay times used for their fit and ours are both $\tau_2 / \tau_1
\approx 10$. Similarly we can make the same observation as made
in~\cite{Greiner} that the momentum distribution width $f_w$ has
returned to its steady state value within a time scale
approximately of order $\tau_{\textrm{tunnel}}$. This is much
shorter than the expected time $t_{\textrm{restore}}$ required for
coherence to spread over the whole lattice of $M$ sites via
single-atom hopping. This confirms that the restoration of
coherence as seen in the experiment is accurately described by the
BHM.

\subsubsection{Validity of the simulation for fast dynamics}

The simulations performed here assume that the dynamics of the
atoms is described by the lowest Bloch band of the optical
lattice. This assumption holds if the typical frequency $f \approx
1/t_{\textrm{ramp}}$ of the ramp in $U$ and $J$ obeys $f \ll \nu$,
where $\nu=\sqrt{4E_RV_0}/2\pi$ is the harmonic approximation of
the excitation frequency to the first excited Bloch
band~\cite{Jaksch98}. The shortest ramping time we considered is
$t_{\textrm{ramp}}=0.1$~ms, while $1/\nu=0.05$~ms for the lattice
on average over the ramping. Because the condition $f \ll \nu$ is
not fulfilled, we numerically calculated the probability of
exciting a single particle, when initially prepared in the lowest
Bloch band and located at the central site of the lattice, during
the ramping above as a function of $t_{\textrm{ramp}}$. We find
that the time evolution is well approximated by the adiabatic time
evolution for $t_{\textrm{ramp}}>0.05$~ms and that it changes to
being sudden for $t_{\textrm{ramp}}<0.005$~ms. Therefore we expect
only a small influence to the form of the curve in
\fir{main_result}(a) between the points at
$t_{\textrm{ramp}}=0$~ms and $t_{\textrm{ramp}}=0.1$~ms due to
higher band excitations which is not resolved in the
experiments~\cite{Greiner}.

\section{Conclusion}
\label{concl}

In these studies we have established the accuracy and
applicability of the TEBD algorithm to the BHM, for both the
computation of ground states and its dynamics. We have then
applied this method to systems of a size equivalent to those
studied in experiments and in the presence of a trapping
potential. In particular we have examined the nature and speed in
which coherence is reestablished within the system for both slow
and rapid dynamics which cross the SF-MI transition. Our results
indicate that for slow ramping of the lattice depth the SF growth
is consistent with single atom hopping as might naively be
expected. However, for very rapid ramping of the lattice depth we
find that the SF growth is much greater than can be explained by
this mechanism alone and so points to the importance of
higher-order correlation functions. We made direct comparisons
between our simulation results for the momentum distribution width
$f_w$ during rapid ramping and the experimental results obtained
by Greiner {\em et al.}~\cite{Greiner} and found that the
reduction in $f_w$ with the ramping time follows precisely the
same functional form as their data, despite a number of
significant differences in the systems analyzed. Perhaps most
fundamentally we have shown that the results obtained
in~\cite{Greiner} for the rapid restoration of coherence are
consistent and explicable within the BHM alone and are present
even in 1D systems. Finally, we note that a detailed knowledge of
the correlations of atoms in different sites and the particle
number fluctuations as provided by our numerical calculations are
important for utilizing the MI state in a number of
applications~\cite{Jaksch99,Bloch,Dorner,Knight,Brennen,Jaksch00,Briegel,Jane,Molmer}.

\acknowledgements D.J. acknowledges useful discussions with
Guifr{\'e} Vidal and the hospitality of Caltech. This work was
supported by EPSRC (UK).

\appendix

\section{1D Optical lattices}
\label{1dol}

The starting point for our physical model is the Hamiltonian for
weakly interacting bosonic atoms in an external trapping potential
\begin{eqnarray}
H &=&\int d{\bf r}\psi ^{\dagger }({\bf r})\left(-\frac{1}{2m_A}
\nabla^{2}+V_{0}({\bf r})+V_{T}({\bf r})\right) \psi ({\bf r})  \label{H} \nonumber \\
&&+\frac{1}{2} \frac{4\pi  a_{s}} {m_A} \int d{\bf r}\psi
^{\dagger }({\bf r})\psi^{\dagger } ({\bf r}) \psi ({\bf r})\psi
({\bf r}),
\end{eqnarray}
with $\psi \left({\bf r}\right)$ the bosonic field operator for
atoms in a given internal atomic state, $V_{0}({\bf r})$ is the
optical lattice potential, and $V_{T}({\bf r})$ describes a slowly
varying additional external trapping potential such as that
created by magnetic fields. The interaction between the atoms is
modeled by a contact potential with $s$-wave scattering length
$a_{s}$ and $m_A$ is the mass of the atoms.

We assume the optical lattice potential to have the form
$V_{0}({\bf r})=\sum_{j=1}^{3}V_{j0}\sin ^{2}(kr_j)$ with wave
number $k=2\pi /\lambda $ and $\lambda $ the wavelength of the
laser light yielding a lattice period $a=\lambda /2$. The spatial
coordinate is denoted by ${\bf r}=(r_1,r_2,r_3)$. This lattice
potential can be realized by interfering three pairs of
counterpropagating laser beams from three orthogonal directions.
The height of the potential $V_{j0}$ is proportional to the
intensity of the lasers in the $j$ th pair of laser beams. We
assume the intensity to be very large in the $r_2$ and $r_3$
directions so that the atoms do not tunnel in either of these
directions. Hence their motion is restricted to the $r_1 \equiv x$
direction and this optical lattice setup thus allows the creation
of effective 1D systems. The resulting atomic pipelines
\cite{Esslinger,Dorner,Paredes} are well isolated from each other
and we can thus restrict our considerations to just one of them.

The center position of the lattice site $m$ of this 1D system is
given by $x_m=m a$, and so a particle occupying the lowest Bloch
band which is localized at this site is described by the wave
function $\phi_m({\bf r})=w_1(x-x_m) w_2(r_2) w_3(r_3)$, where
$w_j$ are the Wannier functions of the lowest Bloch band
\cite{Jaksch98} in the $j$ th direction. By neglecting all
excitations to higher bands and expanding the bosonic field
operators into the mode functions $\phi_m({\bf r})$ the
Hamiltonian $H$ reduces to the 1D Bose-Hubbard model
\cite{Jaksch98} given in \eqr{BHM} Sec.~\ref{model}~A . The
parameter $U$ of the BHM is given by $U=4\pi a_{s} \int d{\bf r}
|\phi_m({\bf r})|^{4}/m_A$ and corresponds to the strength of the
on-site repulsion of two atoms occupying the same lattice site
$m$. The hopping matrix element $J$ between adjacent sites $m$ and
$m+1$ is given by
\begin{eqnarray}
J &=& -\int dx w_1(x-x_m)\bigg(-\frac{1}{2m_A} \frac{d^2}{d
x^2} \nonumber \\
 & & \qquad + \quad V_0 \sin^{2}(kx) \bigg) w_1(x-x_{m+1}).
\end{eqnarray}
The numerical values for $U$ and $J$ for different depths of the
optical lattice $V_0 \equiv V_{10}$ can be found in
\cite{Jaksch98}.

\section{MF approximation and Gutzwiller ansatz}
\label{mfapp}

The presence of a trapping potential causes the system to become
inhomogeneous allowing the coexistence of spatially separated SF
and MI
regions~\cite{Jaksch98,Bruder,Batrouni,Bergkvist,Kollath,Wessel,Kashurnikov,Pollet}.
Now each lattice site has a local chemical potential $\mu_m$, and
together with the overall $U/2J$ parameter for the system this
gives a phase diagram coordinate ${\bf p}=(2J/U,\mu_m/U)$. The MF
determination of which regime a given lattice site lies in is then
based on where precisely ${\bf p}$ resides in the homogeneous BHM
phase diagram \cite{Fisher,Sachdev,Kashurnikov}.

The MF calculations performed utilized the standard decoupling of
the hopping term \cite{Sheshadri,Stoof}
\begin{equation}
b_m^{\dagger }b_{m+1}= b_m^{\dagger }\Av{b_{m+1}} +
\Av{b_m^{\dagger }}b_{m+1} - \Av{b_m^{\dagger }}\Av{b_{m+1}},
\label{MF}
\end{equation}
which decompose the BHM Hamiltonian in \eqr{BHM} into a set of $M$
single site Hamiltonians $H_m$. Specifically the single site $H_m$
has the form
\begin{equation}
H_m = -2J(\phi_m b_m^{\dagger } + \textrm{h.c.}) -2|\phi_m|^2 -
\mu_m b_m^{\dagger }b_{m} + \frac{U}{2} b_m^\dagger b_m^\dagger
b_m b_m, \label{Hm}
\end{equation}
for each lattice site $m$, dependent on the sites superfluid order
parameter $\phi_m = \Av{b_{m}}$, which is assumed to vary slowly
over the system, and its local chemical potential $\mu_m$. The MF
ground state is then a product state over sites
$\ket{\psi_{\textrm{MF}}}=\prod_{m=1}^M \ket{\psi_{0}^{m}}$
determined by minimizing the complete set of $M$-site Hamiltonians
$H_m$ with respects to the order parameters $\phi_m$ and then
diagonalizing to extract eigenstate $\ket{\psi_{0}^{m}}$ with the
smallest eigenvalue for each site. Given that each
$\ket{\psi_{0}^{m}}=\sum_{n_m} c_{n_m} \ket{n_m}$ this procedure
is equivalent to approximation methods based on the Gutzwiller
ansatz \cite{Fisher,Krauth} and has been applied successfully in
modeling the qualitative behavior of BHM phase diagram
\cite{Sachdev}.


\begin{references}
\bibitem{timescale}
We define the timescale $\tau_{\textrm{tunnel}}= \pi/2J$. This is
physically motivated by considering the time required for one
atom, within a double well potential, to hop from one well to
another, given a tunnelling matrix element $J$.
\bibitem{jchoice}
To define a timescale for hopping across half the lattice, when
$J$ varies in time, some appropriate value must be chosen to
represent its typical value during the ramping.
\end{references}
\end{document}